\input psfig
\documentstyle[twocolumn,aps]{revtex}
\begin{document}
\draft

%
\twocolumn[\hsize\textwidth\columnwidth\hsize\csname@twocolumnfalse\endcsname
%
%

\title{Indications of a Metallic Antiferromagnetic Phase in the 2
Dimensional $U-t-t'$ Model}

\author{ Daniel Duffy and Adriana Moreo}

\address{Department of Physics, National High Magnetic Field Lab and MARTECH,
Florida State University, Tallahassee, FL 32306, USA}

\date{\today}
\maketitle

\begin{abstract}
We present mean-field and quantum Monte Carlo results that suggest the
existence of an itinerant antiferromagnetic ground state 
in the half-filled $U-t-t'$ model in two dimensions. In
particular, working at $t'/t=-0.2$ we found that antiferromagnetic
long range order develops at $U_{c_1}/t\approx 2.5 \pm 0.5$, while a study of the
density of states $N(\omega)$ and the response to an external magnetic field
indicates that the system becomes insulating at a larger coupling 
$4<U_{c_2}/t<6$.
\end{abstract}

\pacs{PACS numbers: 71.30.+h, 75.10.Lp, 75.50.Ee}
\vskip2pc]
\narrowtext

%
%
The interest in metal-insulator transitions started several decades ago
with the observation that nickel oxide (NiO), a transparent non-metal,
should be metallic according to its electronic band
structure.\cite{Mott} 
Afterwards, several models were proposed to
study metal-insulator transitions, including the well-known Hubbard
model. Brinkman and Rice found that when the on-site Coulomb repulsion
$U$ reaches $U_c=1.15 W$, where $W$ is the
electronic bandwidth,\cite{BR} the ground state should change from metallic to
insulating.
However, a variety of numerical and analytical studies have convincingly
shown that the half-filled Hubbard model with electronic hopping between
nearest neighbor sites has an insulating
antiferromagnetic ground state for any finite value of the coupling $U$. The
addition of a hopping along the plaquette diagonals to the Hubbard 
Hamiltonian destroys the
nesting, and the possibility of a finite critical coupling ($U_c$) 
at half-filling is
recovered. $U_c$ depends on the diagonal hopping $t'$,
and in previous work 
it has been obtained by monitoring the value of $U$ at which 
antiferromagnetic long range order
(AFLRO) develops.\cite{Hirsch} Mean field and quantum 
Monte Carlo methods have been the techniques most used for this purpose. 
It is generally accepted that the half-filled $U-t-t'$ model has a metallic
ground state for $U<U_c$ and an antiferromagnetic insulating (AFI) state for
$U>U_c$. 

The aim of our work is the study of the metal-insulator transition
(MIT) in the $U-t-t'$ model using updated numerical techniques. We will
monitor the development of AFLRO from the behavior of the spin
correlations, and we will search for the onset of an insulating phase by
studying the density of states and the response
of the system to magnetic fields. The main result is that we have 
found indications
of the existence of an intermediate phase between the paramagnetic metal
and the AF insulator. The phase can be characterized as an AF metal (AFM).
The existence of two dimensional systems with non-ordered
metallic, AFI and AFM phases has recently been discussed.\cite{Fuku}

The $U-t-t'$ Hamiltonian is given by
$$
{\rm H=
-t\sum_{<{\bf{ij}}>,\sigma}(c^{\dagger}_{{\bf{i}},\sigma}
c_{{\bf{j}},\sigma}+h.c.)
-t'\sum_{<{\bf{in}}>,\sigma}(c^{\dagger}_{{\bf{i}},\sigma}
c_{{\bf{n}},\sigma}+h.c.)}
$$

$$
+U{\rm \sum_{{\bf{i}}}(n_{{\bf{i}} \uparrow}-1/2)( n_{{\bf{i}}
\downarrow}-1/2)+\mu\sum_{{\bf{i}},\sigma}n_{{\bf{i}}\sigma} },
\eqno(1)
$$
\noindent where ${\rm c^{\dagger}_{{\bf{i}},\sigma} }$ creates an electron at
site ${\rm {\bf i } }$
with spin projection $\sigma$, ${\rm n_{{\bf{i}}\sigma} }$ is the number
operator, the sum
${\rm \langle {\bf{ij}} \rangle }$ 
runs over pairs of  nearest neighbor lattice sites, and the sum
${\rm \langle {\bf{in}} \rangle }$ 
runs over pairs of lattice sites along the plaquette
diagonals. $U$ is the
on-site Coulombic repulsion, ${t}$ the nearest neighbor hopping
amplitude, ${t'}$ the hopping amplitude along the plaquette diagonals, 
and $\mu$ is the
chemical potential. In the following $t=1$ will be used.

As a first step, the Hamiltonian will be studied using the spin density wave
(SDW) mean-field (MF) approximation.\cite{Bob} 
Proposing as an Ansatz an AF ground state and following
the standard Bogoliuvov 
procedure to diagonalize exactly the resulting MF Hamiltonian 
we found two energy 
bands given by
$$
E_{\bf{k}}^{\pm}=E_{\bf{k}}^d- \mu \pm E_{\bf{k}}^0,
\eqno(2)
$$
\noindent where
$$
E_{\bf{k}}^d=-4 t' cos k_x cos k_y,
\eqno(3)
$$
$$
E^0_{\bf{k}}=\sqrt{\epsilon_{\bf{k}}^2+\Delta^2},
\eqno(4)
$$
\noindent and
$$
\epsilon_{\bf{k}}=-2 t (cos k_x+cos k_y).
\eqno(5)
$$
\noindent $\Delta$ is the MF parameter that, when finite, indicates that the
ground state has AFLRO. Both
$\Delta$ and $\mu$ are obtained by solving the self-consistent
equations:

\begin{figure}[t]
\centerline{\psfig{figure=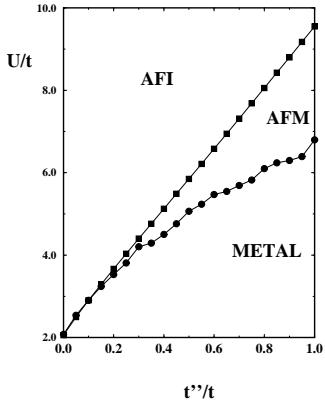,height=5.5cm,bbllx=98pt,bblly=36pt,bburx=590pt,bbury=700pt}}
\caption{SDW mean-field phase diagram of the $U-t-t'-t''$ model at
half-filling and $T=0$, for $t=1$ and $t'=-0.2$.}
\label{fig1}
\end{figure}

$$
{1\over{U}}={1\over{N}}\sum_{\bf{k}}{[f(E^-_{\bf{k}})-f(E^+_{\bf{k}})]\over
{E^0_{\bf{k}}}},
\eqno(6)
$$
\noindent and 
$$
\langle n\rangle= \sum_{{\bf{k}}} n({\bf{k}}),
\eqno(7)
$$
\noindent where $f(x)$ is the Fermi function given by ${1\over{e^{\beta
x}+1}}$, $N$ is the number of sites in the lattice and
$$
n({\bf{k}})={1\over{2}}(1-{\epsilon_{\bf{k}}\over{E^0_{\bf{k}}}})f(E^-_{\bf{k}})+
{1\over{2}}(1+{\epsilon_{\bf{k}}\over{E^0_{\bf{k}}}})f(E^+_{\bf{k}}).
\eqno(8)
$$

Eqs.(6-8) can be solved at any temperature $T$ but since we are trying
to study ground state properties we will work at $T=0$.  To study the
MIT discussed before $\langle n\rangle=1$ is fixed in our calculations.
When $t'=0$ it is found that as soon as $\Delta$ becomes different from
zero, indicating the existence of an AF ground state, a gap of size $2
\Delta$ opens between the two bands given by Eq.(2), and the chemical
potential lies inside the gap. However, when $t'\neq 0$ the shape of the
bands is distorted, and even when AF has developed, the two bands
overlap if $\Delta\leq 2|t'|$. This occurs because the actual gap is
defined by the separation between the highest state in the lower band
and the lowest state in the upper band. For $t'<0$ these states have
${\bf{k}}=(\pi/2,\pi/2)$ and ${\bf{k}}=(\pi,0)$, respectively (for
$t'>0$ the momenta are reversed). For example, for $t'<0$, the lowest
state in the upper band has energy $\Delta-4|t'|$ while the highest
state in the lower band has energy $-\Delta$. An analogous condition is
obtained for $t'>0$. Then, an effective finite gap exists if
$\Delta>2|t'|$.  Therefore, in the region where the bands overlap the
ground state is antiferromagnetic, but it has metallic properties
because the chemical potential cuts both the lower and upper bands.
Unfortunately, solving the MF equations we found that the AFM is not
stable in this case since $\Delta$ changes $discontinuously$ from zero
to $2|t'|$ at a particular value of $U$ which depends on $t'$.  This is
due to the fact

\begin{figure}[t]
\centerline{\psfig{figure=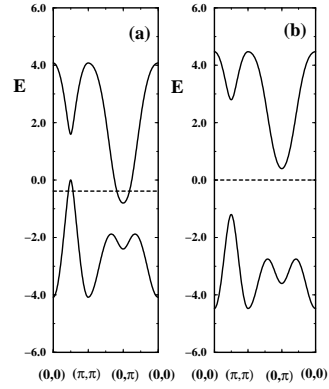,height=5.5cm,bbllx=98pt,bblly=36pt,bburx=590pt,bbury=700pt}}
\caption{SDW mean-field band structure of the half-filled $U-t-t'-t''$
model for $t=1$, $t'=-0.2$ and $t''=0.2$, a) in the AFM phase with $U=3.57$ 
and $\Delta=0.8$ and b) in the AFI phase with $U=5.1$ and $\Delta=2.0$.
The dashed line indicates the position of the chemical potential.}
\label{fig2}
\end{figure}

\noindent that the energy as a function of $\Delta$, in this case, has a
two minima structure.  Then, the AFM phase exists only at one point in
parameter space.  However, it is reasonable to expect that the AFM phase
might be stabilized by including the effect of fluctuations beyond the
MF approximation.  Rather than adding these fluctuations, we intuitively
believe that their effect could be mimicked by introducing longer range
hopping terms in the kinetic energy. Thus, we have added to the MF
calculation an additional hopping term between second nearest neighbors
in the $x$ and $y$ directions with strength regulated by the parameter
$t''$. The MF equations are modified simply by replacing $E_{\bf{k}}^d$
in Eq.(3) by

$$
E_{\bf{k}}^d=-4 t' cos k_x cos k_y - 2 t'' (cos 2k_x+ cos 2k_y).
\eqno(10)
$$

\noindent In this case we found that the overlap of the bands occurs if
$0\leq\Delta\leq 2|t'|+4 t''$. The introduction of $t''$ stabilizes the
AFM phase in the MF approximation as can be seen in the phase diagram
presented in Fig.1 for $t'=-0.2$. The circles indicate the values of
$U_{c_1}$ where the ground state changes from paramagnetic metal to AF
metal, i.e. where $\Delta$ becomes non-zero but the bands still overlap.
The squares represent a second critical coupling, $U_{c_2}$, at which
the ground state becomes an AF insulator. This occurs when $\Delta$
becomes larger than $2|t'|+4t''$ and the bands no longer overlap. In
Fig.2.a the energy bands along certain directions in momentum space in
the AFM phase are shown. As it was discussed above, it is clear that the
two bands overlap and the position of the chemical potential, denoted
with a dashed line, indicates that the top of the lower band is empty
while the bottom of the upper band is filled. The energy bands in the
AFI phase are presented in Fig.2.b. In this case, $\mu$ lies in the
middle of the gap and the system is clearly an insulator.  Then, at the
MF level, the addition of longer range hopping terms stabilizes the AFM
ground state.\cite{foot2}

To find further support for the existence of the AFM 

\begin{figure}[t]
\centerline{\psfig{figure=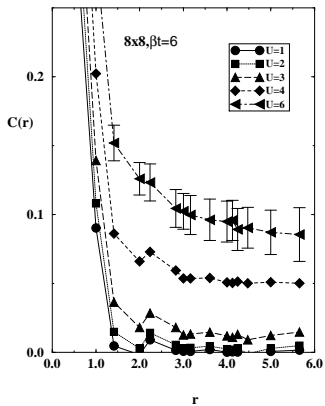,height=5.5cm,bbllx=98pt,bblly=36pt,bburx=590pt,bbury=700pt}}
\caption{QMC spin-spin correlation ${\rm C({\bf r})=\langle S^z_{\bf i}
S^z_{\bf i+r}\rangle (-1)^{|{\bf r}|} }$ for several values of $U/t$,
${\rm T=t/6}$ on an $8\times 8$ lattice at half-filling for $t'/t=-0.2$.
Points without error bars have errors smaller than the size of the dots.}
\label{fig3}
\end{figure}

\noindent ground state, Quantum Monte Carlo (QMC) techniques have been
here applied to the study of the $U-t-t'$ model, working at $t'=-0.2$ on
$8 \times 8$ lattices and temperatures $T=1/8$ and $1/6$.  For these
values of the parameters the MF approximation (with $t''=0$) predicts
$U_{c}=2.1$.  Previous Quantum Monte Carlo results suggested that AFLRO
develops for $U_{c_1}\approx2.5$.\cite{Hirsch} We have independently
analyzed the spin-spin correlation function $C(r)$ as a function of
distance for several values of $U/t$ on an $8 \times 8$ cluster. The
results are shown in Fig.3.  Note that a finite tail is already
developed for $U/t=3$ but it is not present for $U/t=2$. Thus, a
critical coupling $U_c \approx 2.5 \pm 0.5$ is here estimated.\cite{com}
Now let us focus our attention to the coupling $U=4$ where AFLRO is
clearly developed (Fig.3).  The next issue is whether the system is
metallic or insulating for $U=4$. To investigate this important point,
the density of states $N(\omega)$ was calculated using the maximum
entropy technique.\cite{silver} The stability of the results was checked
by making four independent long runs for each set of parameters. In
Fig.4.a the results for $U=4$, $t'=-0.2$ and $\beta=6$ are presented.
Three main peaks are observed in the results: those to the left and to
the right of $\mu$ can be identified with the lower and upper Hubbard
bands, while the peak at $\mu$ corresponds to the quasiparticle weight
indicating that the system is metallic. This metallic behavior
disappears as $U$ increases and an effective gap develops. In Fig.4.b
$N(\omega)$ for $U=6$, $t'=-0.2$ and $\beta=4$ is shown. Here the finite
temperature precursor of an insulating gap in the density of states has
developed and the chemical potential lies inside the gap. Notice that
these results are in agreement with those we obtained in Ref.\cite{DA}
where the gap was studied by monitoring the behavior of the density
$\langle n \rangle$ versus $\mu$, as well as the spectral functions
$A({\bf{k}},\omega)$. In Fig.4.c the mean field result for $N(\omega)$
in the AFM region is presented.  It is remarkable that the three main
peak structure is qualitatively similar to that found numerically for
$U=4$. The MF density of states in the AFI phase is presented in Fig.4.d
where there is also good 

\begin{figure}[t]
\centerline{\psfig{figure=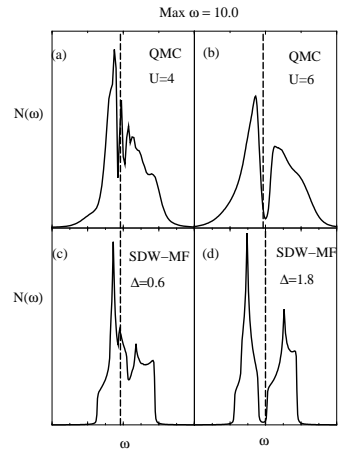,height=5.5cm,bbllx=98pt,bblly=36pt,bburx=590pt,bbury=700pt}}
\caption{QMC density of states $N(\omega)$ for the $U-t-t'$ model at
half-filling for $t'=-0.2$ and a) $U=4$, $\beta=6$, on an $8 \times 8$
lattice: b) same as (a) for $U=6$ and $\beta=4$; c) mean-field results
on a $200 \times 200$ lattice at T=0 using $t''=0.2$, $\Delta=0.6$ and
$\mu=-0.382$; d) same as (c) with $\Delta=1.8$ and $\mu=0$. The MF
transition between AFM and AFI occurs for $\Delta=1.2$}
\label{fig4}
\end{figure}

\noindent qualitative agreement with the QMC numerical results for
$U=6$.

In spite of the good agreement between the SDW-MF and QMC numerical data
suggesting the existence of an AFM phase and a metal-insulator
transition somewhere in the interval $4<U_{c_2}<6$, we decided to study
the lower limit for $U_{c_2}$ using an alternative technique. It is well
known that when a metallic system is close to its critical coupling
towards an insulator, a magnetic field can induce a metal-insulator
transition.\cite{vollhardt,georges} The signature of the transition at
low temperature is a discontinuity (``metamagnetic transition'') in the
magnetization $m$ as a function of the magnetic field $h$ for $U$
slightly smaller than $U_{c_2}$. For a coupling close to but smaller
than $U_{c_2}$ the system is still metallic at zero magnetic field. The
magnetization curve follows a metallic behavior until the field reaches
a critical value $h_c$ that drives the system to become an insulator
and, thus, producing the metamagnetic transition.  Since we are working
at small but finite temperature, we do not expect to find a perfect
discontinuity but instead a very rapid crossover.  In fact, in
Ref.\cite{georges}, where the Hubbard model in infinite dimension was
studied, it was shown that only for $T<0.01$ a proper first order phase
transition is observed.  However, defining $\chi=dm/dh$ as the magnetic
susceptibility, at low but finite temperatures and at low magnetic
fields we expect $d \chi/dh>0$ to be the signature of the metamagnetic
transition that should occur at lower temperature. On the other hand,
$d\chi/dh\leq 0$ would indicate normal behavior.\cite{georges} For a
standard Fermi liquid, $m$ vs $h$ should be linear for small magnetic
fields (i.e., $d \chi/dh=0$). With increasing $h$, the slope will
decrease as $m$ saturates to 1. The numerical calculation of $m$ as a
function of $h$ is very difficult because the behavior that we want to
study occurs at very small magnetic field, i.e. $h< 0.1$. We found that
the anomalous behavior is not observed for temperatures higher than
$T=0.125$. To obtain numerical values

\begin{figure}[t]
\centerline{\psfig{figure=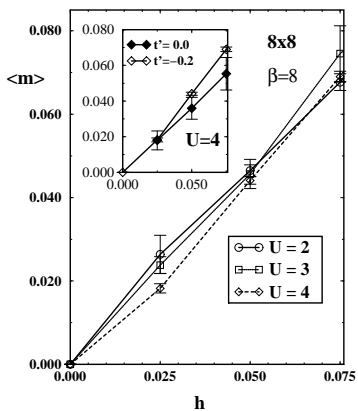,height=5.5cm,bbllx=98pt,bblly=36pt,bburx=590pt,bbury=700pt}}
\caption{Magnetization $m$ versus magnetic field $h$ for the half-filled
$U-t-t'$ model. $t'=-0.2$ and $U=2$, 3 and 4 at T=t/8. In the inset $m$
versus $h$ is shown for $U=4$ and $t'=0$ (filled diamonds) and -0.2
(open diamonds).}
\label{fig5}
\end{figure}

\noindent with small enough error bars we needed to perform about
100,000 measuring QMC sweeps per point.  In Fig.5, $m$ vs $h$ for
$t'=-0.2$ is shown at different values of $U$ for $\beta=8$ on an $8
\times 8$ lattice.  $d\chi/dh\leq 0$ is observed for $U=2$ and 3.
However, for $U=4$ $d \chi/dh>0$ is found. This suggests that at
$t'=-0.2$, $U=4$ is a lower bound for $U_{c_2}$. It could be argued that
a quantum antiferromagnet will have $d \chi/dh>0$ because its
magnetization is given by $m=\chi h+$ sgn $h$ $h^2/4 \pi
c^2$\cite{fischer}. However, for the small values of $h$ used here the
linear term prevails as can be seen in the inset of Fig.5 where the
magnetization as a function of the magnetic field is shown for $U=4$ and
$t'=0$. Since antiferromagnetism is reduced by the addition of a finite
$t'$ the upturn observed for $t'=-0.2$ can not be caused by staggered
spin correlations.  Then, by comparing the two curves in the inset of
Fig.5, it is clear that $d \chi/dh$ becomes larger than zero due to the
effect of the diagonal hopping. It appears that in the half-filled
$U-t-t'$ model when $t'=-0.2$, AFLRO develops at $U_{c_1}=2.5 \pm 0.5$
but the system becomes an insulator for $U_{c_2}$ larger than 4. These
numerical results again suggest that the AFM region exists, and it is
actually broader than what was predicted at the mean field level. Notice
that the existence of AFM phases is not just an academic curiosity.
Magnetically ordered metallic phases have been observed
experimentally.\cite{Bao} In particular, an AFM phase was found in
organic $\kappa-({\rm{BEDT-TTF}})_2 {\rm{X}}$.\cite{org}

Summarizing, in this paper we have provided numerical and analytical
results that suggest the existence of an antiferromagnetic metallic
ground state in the two dimensional $U-t-t'$ model.\cite{foot}

We thank V.~Dobrosavljevic, W.~Brenig, E.~Miranda, E.~Dagotto,
D.~Scalapino, S.~Valfells and S.~Hill for useful conversations.  A.M. is
supported by NSF under grant DMR-95-20776. Additional support is
provided by the Office of Naval Research under grant N00014-93-0495, the
National High Magnetic Field Lab and MARTECH.  We thank ONR for
providing access to their Cray-YMP and CM5 supercomputers.


\begin{references}
\bibitem{foot} An additional procedure to show numerically 
that the intermediate
phase is indeed metallic would be by studying the behavior of the Drude
weight that acts as order parameter 
(see  E.~Dagotto, Rev.~Mod.~Phys. {\bf{63}}, 763 (1994) and
references therein). However, the standard
sum rule as well as the definition of the current are modified by the
introduction of the diagonal hopping. Work along this line
is in progress.
\end{references}
\end{document}